# PERFORMANCE AND NEAR-WAKE CHARACTERIZATION OF A TIDAL CURRENT TURBINE IN ELEVATED LEVELS OF FREE STREAM TURBULENCE

**Ashwin Vinod and Arindam Banerjee[1]**

Department of Mechanical Engineering & Mechanics
Lehigh University, Bethlehem, PA

[1]Corresponding Author:
19 Memorial Drive West, Mechanical Engineering & Mechanics, Lehigh University,
Bethlehem, PA 18015
Phone: 610-758-4099; Email: arb612@lehigh.edu

# ABSTRACT

Tidal turbines are deployed in sites which have elevated levels of free stream turbulence (FST). Accounting for elevated FST on their operation become vital from a design standpoint. Detailed experimental measurements of the dynamic near-wake of a tidal turbine model in elevated FST environments is presented; an active grid turbulence generator developed by our group was used to seed in the elevated FST and evaluate the influence of turbulence intensity ($T_i$) and inflow integral length scale ($L$) on the near-wake of the turbine. Three inflow conditions are tested: a quasi-laminar flow with $Ti \sim 2.2\%$ and two elevated $Ti$ (~12-14%) cases, one with $L \sim 0.4D$ ($D$ is the turbine diameter) and the other where $L \sim D$. Elevated $Ti$ cases was found to increase the standard deviation of rotor torque by 4.5 times the value in quasi-laminar flow. Energy recovery was also found to be accelerated; at $X/D=4$, the percentage of inflow energy recovered was 37% and was twice the corresponding value in quasi-laminar flow. Elevated FST was observed to disrupt the rotational character of the wake; the drop in swirl number ranged between 12% at $X/D=0.5$ to 71% at $X/D=4$. Elevated $Ti$ also resulted in $L$ that were considerably larger (> 2 times) than the quasi-laminar flow case. An increase in inflow integral length scale (from $0.4D$ to $D$) was observed to result in enhanced wake $Ti$, wake structures and anisotropy; however, no noticeable influence was found on the rate of wake recovery.

# 1. INTRODUCTION

Tidal flows across the globe are considered to be a highly predictable renewable energy resource as tides can be predicted weeks, months, or even years in advance [1]. One of the major advantages of tidal stream turbines (TST) in that it can be sized to fit the local environment, i.e. tidal range, tidal flow or coastal restrictions [2]; typical commercial turbines have variable pitch blades with a diameter of 5-25 m with rated power between 35 kW to 2 MW[3]. Depending on the application, TSTs can be placed either as individual deployments for niche markets[4] or in large arrays (farm configurations) to maximize the electricity produced at a given site [3, 5]. There have been several advanced deployments towards installing full-scale arrays or farms of TSTs in the European Union, a review of these arrays can be found elsewhere [6, 7]. In the United States, Verdant Power Inc. is installing an array of tidal turbines as a part of the Roosevelt Island Tidal Energy (RITE) project; a total of ten tri-frames with a three-bladed turbine positioned in a triangular arrangement [8] will be deployed in the East River in New York City in 2020 as a part of a Phase 3 U.S. Department of Energy demonstration project [9]. As in wind farm layouts, the density of the TST distribution within an array is determined by wake recovery as velocity deficit in the wake generated by the upstream TST reduces the power output for the downstream turbine. Several studies have discussed wake hydrodynamics of single as well as arrays of TST; the main focus is to quantify the loading[10-12], blockage [13, 14], wake size[15] and wake meandering under yawed inflow[16-18] or proximity to free-surface [19, 20]. An understanding of the wake size and wake merger as it meanders downstream is of considerable benefit to the design engineer responsible for spacing the TSTs in the array [21].

Tidal flows are considerably complex and turbulent in nature[22]. Turbulent inflow may result in unsteady variations on the angle of attack with respect to the rotor blade that can lead to the occurrence of the dynamic stall and reduce the efficiency of the turbine. In addition, largely increased turbulence intensity in the wake is expected to increase the dynamic loads and hence negatively affect the fatigue life of the turbine blades [23, 24]. One-dimensional turbulence intensity (*Ti*) is defined as

$$Ti = \frac{\sigma(u)}{U} \times 100 \tag{1}$$

where $\sigma(u)$ is the standard deviation of principal velocity samples, and $U$ is the corresponding time-averaged value. *Ti* levels vary from one tidal site to the other, and also within site depending on time and location. Several studies have assessed resource characteristics at tidal sites in the U.S. and U.K.; the findings are summarized in Table 1 below. In most cases, natural tidal flows had mean velocities between 1.3-3.5 m/s with elevated levels of *Ti* in the 8% - 13% range, categorizing the tidal conditions into sub-classes B (10% < *Ti* < 15%) and C (*Ti* < 10%) as per the marine energy classification system [29]. The only noted exception was the East River in New York City, where, sub-class A category (*Ti* > 15%) *Ti* levels in the 20 - 30% range were reported. There have been only a few studies that have explored the effects of elevated free-stream turbulence (FST) levels on the performance, loading, and wake characteristics of a tidal current turbine. Blackmore *et al*. [24] investigated the effect of turbulence on the drag of solid and porous disc turbine simulators and found the drag coefficient to be considerably affected by the level of inflow turbulence. In some cases, the drag coefficient measured in flows with higher *Ti* (>13%) was found to be greater by 20% when compared to the cases at *lower Ti* (<4.5%). Blackmore *et a*l.[23] also investigated the effect of elevated turbulence (5-10%) on the wake generated by an actuator disc using large eddy simulations (LES) and predicted that the elevated turbulence would lead to a quicker recovery of the wake due to enhanced mixing. More recently, they performed an experimental study to investigate the effects of turbulence intensity (5%-25%) and integral length scale (0.18*D* to 0.51 *D*, where *D* is the turbine diameter) on a lab scale tidal turbine model [30]. While an increase in turbulence intensity resulted in ~ 10% drop in power and thrust coefficients, increase in integral length scale was observed to result in a ~10% increase in the two non-dimensional coefficients; both inflow parameters resulted in increased load fluctuations on the rotor. They also proposed that the fatigue loads acting on the turbine could be estimated from the fluctuations observed in the power output and could, in turn, be used to optimize maintenance operations without the need for additional monitoring instrumentation. Mycek *et al.* [31, 32] tested a tidal turbine at two different values of *Ti*; 3% and 15% in the IFERMR (French Research Institute for Exploitation of the Sea) wave and current flume tank. In their experiment, the higher *Ti* inflow was attained by removing the flow conditioning screens upstream of the test section; the work did not use a dedicated turbulence generator as such. In addition to turbine performance, they measured its wake characteristics along the axial plane (in the 1.2*D* to 10*D* downstream distance range) at a tip speed ratio that corresponded to its peak performance, however, did not discuss the aspect of integral length scales. Payne et al.[33] studied the frequency characteristics of unsteady loads acting on a bed-mounted, 1:15 scale tidal turbine model, in the same facility at IFERMR. They found a noticeable correlation between the spectra of rotor loads and onset velocity, more prominent at frequencies lower than the rotational frequency of the rotor. They also analyzed the streamwise and transverse forces acting on the turbine support structure and identified two vortex shedding regimes associated with the wake and bypass flow.

| Locations | U (m/s) | Ti (%) |
|---|---|---|
| Fall of Warness, U.K. | 1.5 | 10-11 |
| Sound of Islay, U.K. | 2.0 | 12-13 |
| Puget Sound, WA, U.S. | 1.3(±0.5) | 8.4/11.4 |
| Strangford Narrows | 1.5-3.5 | 4-9 |
| East River, NY, U.S. | 1.5-2.3 | 20-30 |

Table 1. Turbulence Intensities (Ti) measured at potential turbine deployment sites[12, 25-28].

As implied by the existing literature, both the inflow turbulence parameters, turbulence intensity, and integral length scale could have a considerable influence on the performance and wake characteristics of a tidal turbine. Experimental data sets illustrating the effects of inflow turbulence on the characteristics of wake development downstream of a tidal stream turbine model are very limited, and the unexplored parameter space is large. In fact, to the knowledge of the authors, the only available data sets are those reported by Mycek et al. [31]. A thorough understanding of wake evolution in a turbulent flow environment is essential in developing efficient and robust tidal farm designs. The current work provides a detailed experimental illustration of the effects of homogenous FST (generated using an active grid) on the performance and near wake characteristics of a lab-scale tidal turbine. Focusing on the more dynamic near wake region of the turbine, the presented results complement and add to the results reported in Mycek et al. [31]. The paper primarily aims to provide a quantitative description of the characteristics of turbine wake generation and propagation in a well-characterized turbulent inflow. The wake measurements reported in this paper include downstream locations as close as half a rotor diameter, which to the knowledge of the authors, is the nearest experimental characterization of a turbine wake in a turbulent flow. In addition to time-averaged and turbulent velocity characteristics discussed by Mycek et al. [31]. and elsewhere by the authors[34], the presented results also discuss the impact of inflow turbulence on wake rotation and the evolution of integral length scales within the wake. The paper also explores the effects of inflow integral length scale in the unexplored range $0.4D$ to $D$, to decouple the effects of inflow turbulence intensity and inflow integral length scales, thereby identifying aspects of turbine performance and wake characteristics that are sensitive to the two turbulent inflow parameters. To better assess (quantitatively) the implications of inflow turbulence for a tidal farm scenario, the rate of energy recovery within the turbine wake, and the propagation of turbine generated periodicities are also presented and discussed. Though restricted to the near wake region, the current datasets can be used to validate and verify numerical models to better incorporate ambient turbulence effects[35] into their predictions.

## 2. EXPERIMENTAL METHODS

All experiments reported in this paper were conducted in an open surface, recirculating water tunnel (Engineering Laboratory Design Inc., Model# 505), housed at Lehigh University. The facility has a test section that is 0.61 m wide, 0.61 m tall and 1.98 m long. It is equipped with a 25HP single stage axial flow propeller pump and is capable of attaining flow speeds up to 1m/s in the test section.

### 2.1 Tidal Turbine Model and Performance Characteristics

A three-bladed tidal turbine model (1:20 scale) with a diameter ($D$) of 0.2794m, was used for the current experiments [see figure 1(A) for a schematic of the turbine assembly]. The design was developed in house and used in previous studies[19, 34]; the rotor blades are made of corrosion resistant aluminum alloy with an SG6043 hydrofoil cross-section having a constant chord ($c$=0.0165m), pitch angle of $10^0$, and no twist [see figure 1(B) for the blade specifications]. Though a simpler blade profile, the results obtained using the rotor would be applicable to tidal turbines in general. The turbine operates at a blockage ratio of 16.5% in our water tunnel. A stepper motor (Anaheim Automation, Model# 23MDSI) connected to the rotor via a stainless-steel shaft precisely controls its rotational speeds. In all the tests presented in the paper, the turbine was operated in a constant RPM mode. The rotor-shaft-motor assembly is mounted directly on to a load measuring setup, which decouples and measures the thrust and torque acting on the rotor. The thrust sensor (Interface, Model# SML-25) and a torque sensor (Interface, Model# MRT2P) used in the turbine model have a manufacturer prescribed non-repeatability of ±0.0334N and ±0.001Nm respectively.

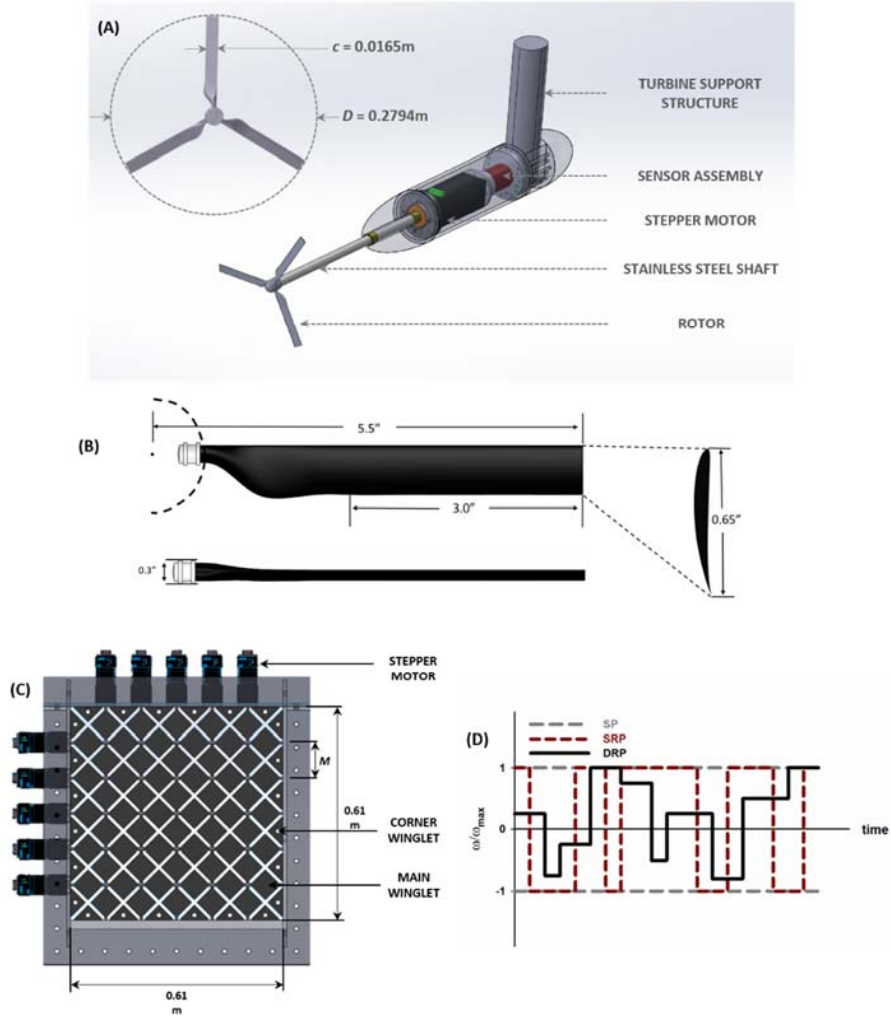

Figure 1. Schematic of (A) laboratory (1:20) scale tidal turbine model, (B) Blade profile (SG-6043) of the turbine rotor, (C) Lehigh active grid, (D) active grid operating protocols ($\omega$ is the instantaneous angular velocity in rad/sec, and $\omega_{max}$ is the prescribed maximum angular velocity in rad/sec)

All electronic components in the turbine assembly are enclosed in a watertight, pressurized (~20kPa gauge pressure) acrylic cylindrical casing. The sensors captured the thrust ($T$) and torque ($Q$) acting on the rotor at a rate of 200 samples/second [36]. A 300 seconds long time trace of $T$ and $Q$ was collected and analyzed to identify an appropriate duration for the sampling time. For each parameter of interest, a sampling period dependent percentage error was defined ($\%Error = 100|X_n - X_m/X_n|$) such that $X_n$ is the value of a parameter calculated from the complete time trace (300 seconds long in this case), and $X_m$ is the value of the same parameter calculated from shorter, *n* seconds long time traces with *n* ranging from 1-300 seconds. The dependence of time-averaged *T* and *Q*, and their respective standard deviations $\sigma(T)$ and $\sigma(Q)$ on the sampling period are shown in figure 2(A). It was observed that *T*, *Q*, and $\sigma(Q)$ fall below a 3% error within 30 seconds of sampling, whereas, $\sigma(T)$ takes a longer period of ~ 60 seconds to drop below the 3% error mark. Therefore, to minimize errors in calculated performance parameters, the loads acting on the rotor were sampled for a period of 120 seconds. The turbine performance was characterized using the non-dimensional parameters, the coefficient of power ($C_P$), the coefficient of thrust ($C_T$) and tip speed ratio (*TSR*) defined below,

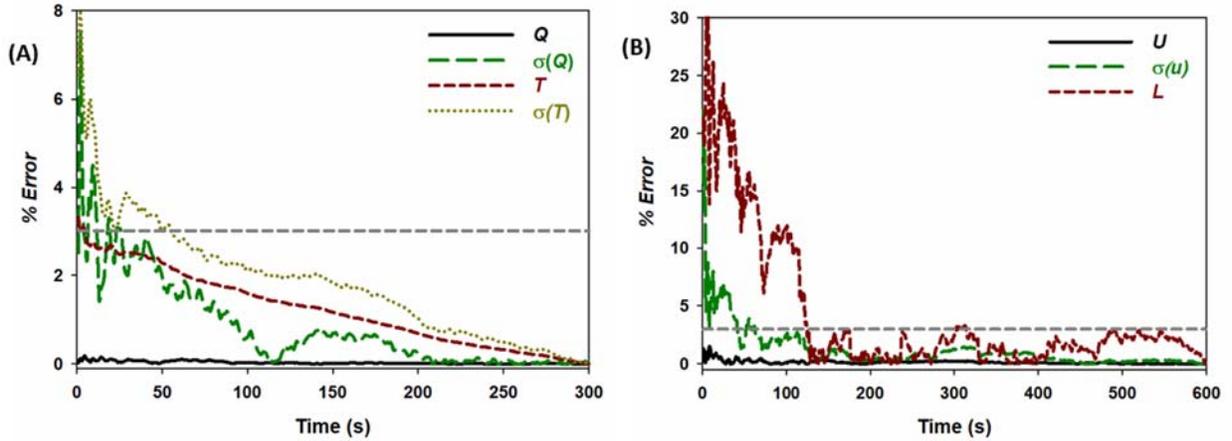

Figure 2: Effect of sampling period on (A) time-averaged thrust, *T*, and torque, *Q*, and their standard deviations $\sigma(T)$ and $\sigma(Q)$, (B) time-averaged velocity *U*, its standard deviation $\sigma(u)$ and the integral length scale, *L*

$$C_P = \frac{Q\Omega}{0.5\rho A U_\infty^3}, \quad C_T = \frac{T}{0.5\rho A U_\infty^2}, \quad TSR = \frac{R\Omega}{U_\infty} \tag{2}$$

where Ω is the angular velocity of the rotor (rad/s), $U_\infty$ is the area averaged freestream velocity (m/s), $\rho$ is the density of water (kg/m$^3$), and $A$ is the area of the rotor (m$^2$). The measured thrust and torque values were corrected for blockage using the method described in Bahaj et al. [37]. To examine Reynolds number (*Re*) dependence, turbine performance was evaluated at 5 values of freestream velocity ($U_\infty$) ranging from 0.5m/s to 1.0 m/s. A considerable change in the $C_P$ envelope was noticed between up to velocities of 0.83m/s. However, the $C_P$ curves obtained for flow velocities greater than 0.83m/s up to a value of 0.95m/s (which corresponds to the maximum flow speeds for the facility) were found to be comparable in terms of max $C_P$ suggesting a reduced dependence on the operating *Re* in this velocity range. Therefore, a free-stream velocity of 0.83m/s was chosen to carry out all experiments, and it corresponds to a diameter-based *Reynolds number* ($Re_D$) of 2.2×10$^5$ and a chord-based *Reynolds number* ($Re_c$) of 1.3×10$^4$. The performance results presented, however, are not claimed to be Reynolds number independent.

## 2.2 Turbulence Generation

Owing to its practical relevance, various mechanisms/designs have been developed to seed turbulence into laboratory flows. Such mechanisms can be broadly classified into two categories; passive generators and dynamic (or active) generators. In the simpler passive generators (also known as passive/static grids), flow turbulence is generated by passing a fluid stream through stationary (static) circular/rectangular rods; the interaction between the wakes formed behind the rods and the jets formed in the spacing between the rods, creates a uniform and transversely homogeneous, decaying turbulent flow. Several experiments involving passive grids have been performed in the last five decades and have been used for the most part in wind tunnels to simulate elevated levels of turbulence for various aerodynamics applications[38]. Compared to static generators, dynamic turbulence generators are more complex; they incorporate moving components into their design that play a significant role in the generated turbulence. Makita and co-workers[39] developed a different type of turbulence generator in Japan in the early '80s for use in their wind tunnel; agitator winglets were attached to rotating rods creating time-varying solidity. This facilitated the production of large scale, high-intensity turbulent flows in small, low-speed facilities producing $Ti$ as high as 16%. Similar active grid designs were later employed by Mydlarski & Warhaft [40] to systematically test the design's capability of generating high Reynolds number turbulence; and by Kang *et al.*[41]to compare LES predictions (of the grid generated turbulence) with the experimental observations. Poorte [42] used an active grid in a vertical test section water tunnel and studied the effects of winglet shaft rotation rates, direction, and their distribution on the turbulence generated by the grid.

The turbulence generator employed in the current work [see figure 1(C)] is a Makita type [39, 43] active grid. It is made of ten rotating winglet shafts; five oriented horizontally and five vertically. Six square winglets, each with a side of 0.06m are attached to the shafts in a diamond fashion (diagonals aligned to the shaft axis), resulting in mesh size, $M$ of 0.1m. The module has an inner cross-section that is identical in size to the test section and is placed at a distance 0.22m (2.2$M$) upstream of the test section entrance. The shafts arranged in a bi-planar configuration are controlled by a dedicated stepper motor, (Anaheim Automation, Model No. 23MDSI) which includes an on-board simple controller/indexer and a micro-stepping driver. In all our tests, the motors were operated at 1600 steps/revolution giving a resolution of 0.225º. In addition to 60 rotating winglets (main winglets), the grid is also fitted with 24 non-rotating, corner winglets, along the inner perimeter of the active grid frame, upstream of the plane of rotating winglets. PTFE V-ring seals are used at the shaft ends closer to motors to minimize water leakage and shaft friction. The three commonly used forcing protocols, synchronous protocol (SP), single random protocol (SRP), and double random protocol (DRP) were programmed using LabVIEW (see figure 1(D) for schematic). In SP, each winglet shaft rotates at a constant angular velocity for the duration of operation with adjacent shafts rotating in opposite directions. It produces lower turbulence intensities, homogeneity, and isotropy. SRP maintains the angular velocity constant while switching the direction of rotation of shafts at random intervals. It produces higher turbulence intensities, better isotropy, and homogeneity when compared to SP. DRP randomizes both the angular velocity and direction of rotation of the shafts and produces higher turbulence intensities and better isotropy and homogeneity[42]. In the current work, primarily two FST conditions were initially selected for studying turbine performance and wake characteristics; the *quasi-laminar flow* case refers to the undisturbed free-stream obtained in the absence of active grid, and the *elevated Ti* case refers to the free-stream disturbed by the active grid operating in DRP, with a maximum winglet shaft angular velocity ($\omega_{max}$) of 3.14 radians/sec. To minimize blockage effects caused by the active grid at the high velocities required for turbine testing, a five-shaft configuration was adopted such that the horizontal shafts were only present and the vertical shafts were removed.

All flow velocity measurements presented in this paper were carried out using a Nortek Vectrino+ Acoustic Doppler Velocimeter (ADV), having a measurement accuracy of ±0.005 m/s. The instrument was set to a sampling frequency of 50Hz[24] and a sampling volume of $1.7 \times 10^{-7}$ m$^3$. The time traces obtained from the experiments were filtered using the phase space thresholding (PST) technique described in Goring and Nikora [44] to eliminate spikes and ensure quality data for further analysis. The spikes identified using the PST algorithm were replaced with the mean value of the time trace. Following Reynolds decomposition, each velocity component $u(t)$ (also $v(t)$ and $w(t)$) can be broken down as follows

$$u(t) = U + u'(t) \tag{3}$$

into a time-averaged component, $U$ and a time-dependent fluctuating component $u'(t)$. The non-dimensional Reynolds stress component $R_{XY}$, anisotropy ratio $I_{XY}$, and the integral length scale $L$, are computed as follows

$$R_{XY} = \frac{\sqrt{\left|\frac{1}{N}\sum_{1}^{N} u_i' v_i'\right|}}{U} \tag{4}$$

$$I_{XY} = \frac{\sigma(u)}{\sigma(v)} \tag{5}$$

$$L = U \int_{0}^{T} \frac{R(dt)}{R(0)} dt \tag{6}$$

with $N$ being the total number of samples, $T$ the corresponding total sampling period and $R(dt)$ the autocovariance defined as

$$R(dt) = \langle u'(t) u'(t+dt) \rangle \tag{7}$$

To ensure the adequate duration of the sampling period, a 600 seconds long time trace was collected and analyzed to assess convergence trends of the different flow parameters of interest. Figure 2(B) shows the sampling period effects on $U$, $\sigma(u)$ and $L$. The %Error in $U$ can be seen to converge almost immediately just within a few seconds; $\sigma(u)$ and $L$ however, took longer to converge and fell below a 3% error mark beyond 60 and 120 seconds respectively. Therefore, a sampling period of 300 seconds was maintained for most of the ADV flow measurements, unless otherwise noted.

The streamwise decay of $Ti$, measured along the center-line of the test section in the quasi-laminar flow case and the elevated $Ti$ case is shown in figure 3(A). In the quasi-laminar flow setting, $Ti$ along the length of the test section remains reasonably constant at 1.9%–2.2%. At the elevated $Ti$ generated using the active grid, $Ti$ can be observed to vary from a value of 24.3% at $X_0/M = 2.8$ to 7.7% at $X_0/M=20$, where, $X_0$ represents the distance downstream of the active grid.

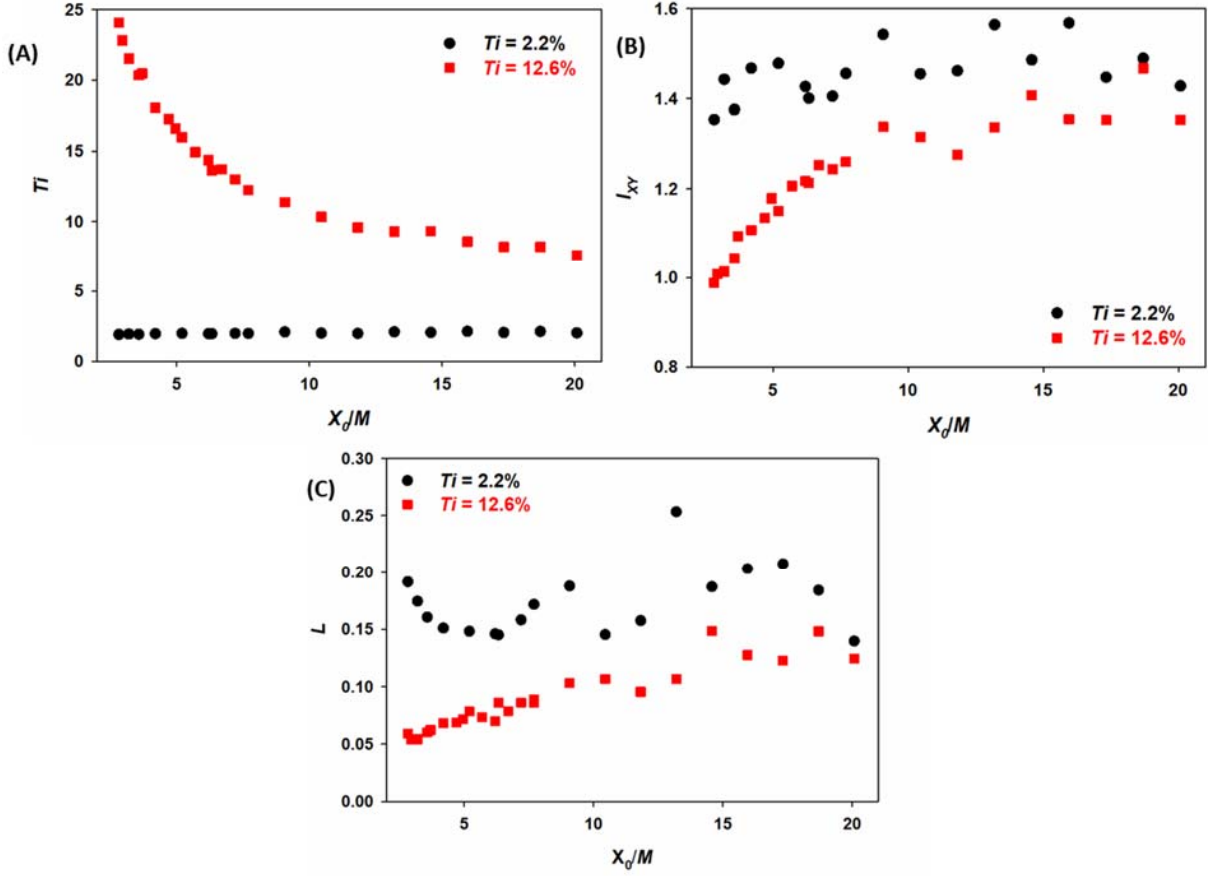

Figure 3: Turbulence characteristics in the tunnel: (A) streamwise decay of $Ti$, (B) streamwise variation of $I_{XY}$, (C) streamwise variation of $L$

The streamwise decay of turbulence in the elevated $Ti$ case was observed to obey a power law of the form

$$\frac{\sigma(u)^2}{U^2} = B\left(\frac{X_0}{M}\right)^{-n} \qquad (8)$$

where the coefficients $B$ and $n$ were estimated to be 0.46 and 1.23, respectively, within the range typically observed for grid turbulence[45]. The streamwise variation of $I_{XY}$ for the two cases is presented in figure 3(B). In the quasi-laminar flow, $I_{XY}$ along the test section was found to remain fairly consistent within 1.46 ± 0.1. At elevated $Ti$, on the other hand, $I_{XY}$ was found to vary from being near isotropic ($I_{XY}$ =1) at $X_0/M$ =2.8 to ~40% anisotropy beyond $X_0/M$ = 15. Figure 3(C) illustrates the downstream evolution of the integral length scale, $L$. A considerable scatter is noticeable in the $L$ estimates obtained in the quasi-laminar flow, mostly varying between 0.15m – 0.20m. In the elevated $Ti$ case, the expected gradual increase in $L$ with downstream distance is noticeable, varying from a value ~0.05m at $X_0/M$ =2.8 to values > 0.10m beyond $X_0/M$ = 15.

### 2.3 Inflow and Wake Characterization

In all experiments with the turbine, the rotor was placed at a downstream distance of $2D$ from the test section inlet (see figure 4(A)) in order to avail considerably high levels of inflow turbulence on the rotor plane without compromising flow homogeneity (will be discussed next in this section). In addition, this configuration also ensured a sufficient test section length downstream of the rotor to capture near wake characteristics. Based on a reference frame ($XYZ$) with origin at the rotor hub ($X_0 = 2.8D$), flow characterization was performed at locations $X/D = 0$, 0.5, 1, 2 and 4. At the downstream locations, $X/D = 0.5 – 4$ in the near wake of the turbine, flow measurements were made at a total of 279 points (referred to as the interrogation region) in the $YZ$ plane as depicted in figure 4(B). Depending on the downstream location, certain points close to the turbine shaft/ nacelle were inaccessible for measurement and hence were excluded from the study. At $X/D = 0$, a coarser measurement grid with a total of 153 points (all adjacent points separated by $0.0909D$ (0.0254m)) was used to characterize the inflow in the absence of the turbine. The different locations within the measurement grid were sampled for a period of 60 seconds, except the locations along $Y=0$ and $Z=0$, which were sampled for a period of 300 seconds for better confidence in the higher order statistics calculated.

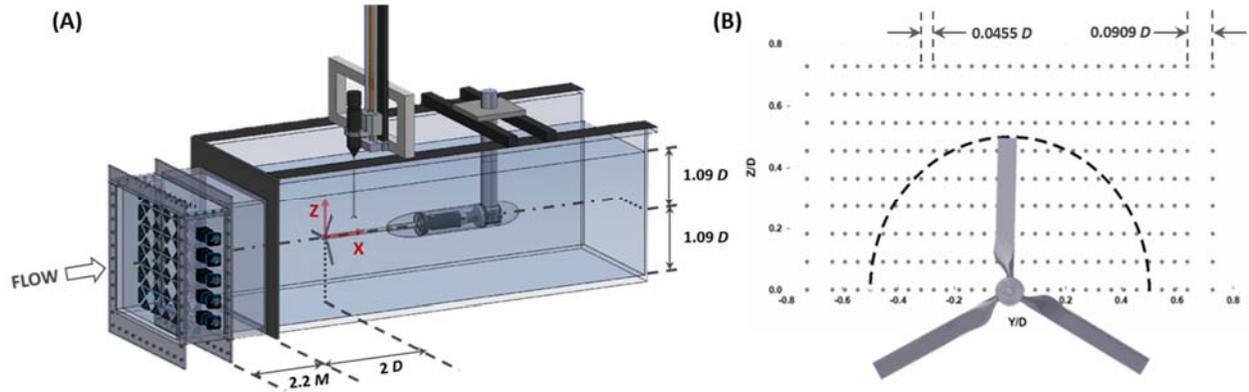

Figure 4: (A) Schematic of the test setup used for the tidal turbine model, (B) Flow measurement locations downstream of the turbine

Figure 5 (A,B,C,D) shows the $U$ and $Ti$ distribution across the interrogation region at $X/D=0$. The average turbulence intensity across the interrogation region was found to be 2.2% in the quasi-laminar flow case and 12.6% in the elevated $Ti$ case; the averaged inflow velocity was found to be 0.83 m/s in the quasi-laminar flow and 0.85 m/s at the elevated $Ti$  The non-uniformity in flow velocity was mostly restricted to values less than 3% of the area averaged velocity in the quasi-laminar flow case and ~ 6% in elevated $Ti$ case. A tendency of the flow to accelerate closer to the free surface was observable at the elevated $Ti$. Table 2 lists all the important parameters used concerning the quasi-laminar and elevated $Ti$ cases. From the table, it can be seen that, in addition to the variation in $Ti$ levels, the integral length scale, $L$, for the two cases are notably different. To decouple the effects of the two inflow parameters, a third free-stream condition, with a $Ti$ comparable to the elevated $Ti$ case and a considerably larger $L$, on the order of the rotor diameter, referred to as elevated $Ti$-$L_D$ was also generated. DRP with a maximum angular velocity ($\omega_{max}$) of 0.785 radians/sec was used for this purpose. The $U$, $Ti$ variation across the interrogation region, and the important velocity statistics corresponding to the elevated $Ti$-$L_D$ are included in figure 5 (E,F) and Table 2. It can be seen that the elevated $Ti$-$L_D$ case has an area-averaged $Ti$ that is ~10% larger than the elevated $Ti$ case, and an area-averaged $L$ that is 2.5 times larger. As in the elevated

*Ti* case, the non-uniformity in mean velocity in the elevated *Ti*-L$_D$ case was mostly restricted to values under 6%. Therefore, a comparison between turbine performance/ wake characteristics when operated in the elevated *Ti* and elevated *Ti*-L$_D$ conditions would highlight the effect of the integral length scale of the inflow.

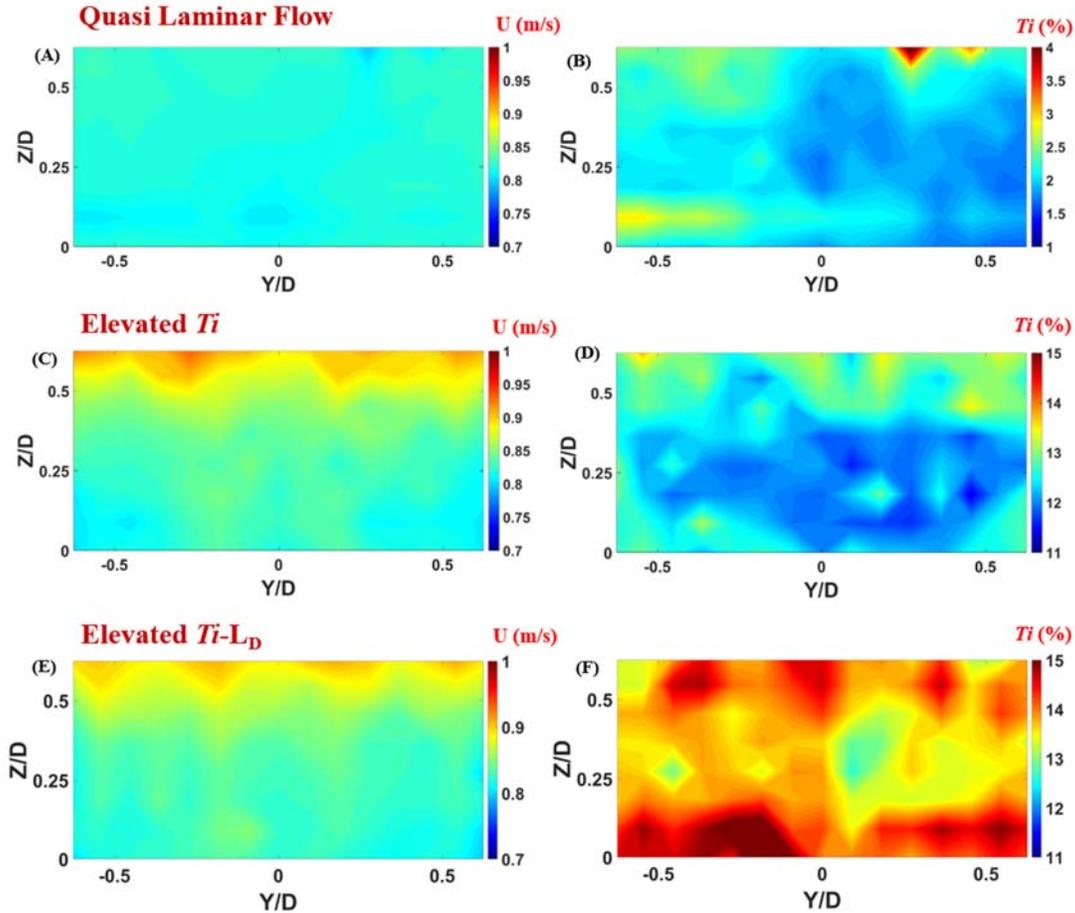

Figure 5. (A), (C), (E) Mean velocity, (B), (D), (F) streamwise *Ti* across the interrogation region for the three inflow conditions

|  | **Quasi-Laminar Flow** | | **Elevated *Ti*** | | **Elevated *Ti*-L$_D$** | |
|---|---|---|---|---|---|---|
| *Parameter* | *At Center-point Value* | *Spatially Averaged* | *At Center-point Value* | *Spatially Averaged* | *At Center-point Value* | *Spatially Averaged* |
| $U_\infty$ | 0.82m/s | 0.83 m/s | 0.82 m/s | 0.85 m/s | 0.82 m/s | 0.85 m/s |
| *Ti* | 1.8% | 2.2% | 11.9% | 12.6% | 13.9 | 13.9% |
| *L* | 0.7*D*/11.2*c* | 0.8*D*/13.6*c* | 0.3*D*/5*c* | 0.4*D*/6.9*c* | 0.9*D*/15*c* | *D*/17.5*c* |
| $I_{XY}$ | 1.13 | 1.3 | 1.27 | 1.3 | 1.43 | 1.4 |

Table 2. Velocity statistics for quasi-laminar, elevated *Ti* and elevated *Ti*-L$_D$ inflows

The power spectral density (*PSD*) for the three inflow cases, measured at the turbine center is shown in figure 6 as a function of frequency, *f*. The differences in the energy content of the different scales in the flow are evident in the figure; the elevated *Ti* cases have a higher energy content across the frequency space and display the -5/3 slope characteristic of the inertial subrange. The presence of larger length scales in the elevated *Ti*-$L_D$ case is clearly reflected in the higher energy content exhibited by, the lower frequencies, and as a result, a longer inertial subrange can be noticed. At the higher frequencies, a minor deviation in the slope is noticeable and is expected to be a consequence of interference with the noise level of the instrument. In the following results section, the effect of inflow turbulence intensity is discussed first in sections 3.1 to 3.7 by comparing the variations in turbine performance and wake characteristics in the quasi-laminar and elevated *Ti* cases. The effects of inflow integral length scales are discussed in section 3.8 using the elevated *Ti* and elevated *Ti*-$L_D$ flows generated.

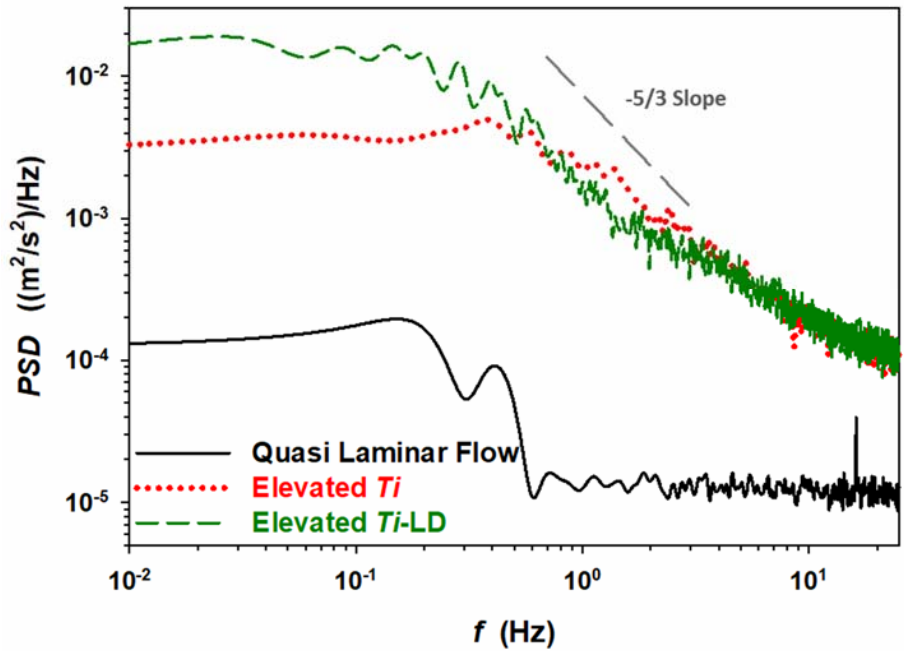

Figure 6. Power Spectral Density (PSD) vs. frequency (*f*) at the three inflow conditions

## 3. RESULTS AND DISCUSSION

### 3.1 Effect of elevated turbulence on Turbine Performance

The performance characteristics of the turbine were assessed for both *Ti*. Similar to the observations reported by Mycek et al.[31], the time-averaged values of $C_P$ and $C_T$ were not affected significantly by a six-fold increase in the level of inflow turbulence (see figure 7 (A)-(B)). The $C_P$ curves can be observed to be almost overlapping at values of $TSR \leq 3$. However, beyond a $TSR$ of 3, the effect of inflow turbulence became more noticeable. In elevated *Ti*, a <10% drop in $C_P$ was observed within $3 \leq TSR < 5$; at $TSR > 5$, $C_P$ dropped by about ~12-13%. $C_P$ attained its maximum value within 4.5 < *TSR* < 5 for both turbulence intensities. $C_T$ was, however, found to be sensitive to the increase in turbulence intensity across all *TSR*'s tested. For the elevated *Ti* case, a drop in $C_T$ of up to 18% was observed at *TSR* ≤ 3; the drop was however restricted to values < 10% at *TSR* > 3. The effect of inflow turbulence was more pronounced on the load fluctuations acting on the turbine[33]. The standard deviations of $C_P$ and $C_T$ are shown in figure 7 (C)-(D), display increased sensitivity to inflow turbulence at *TSR* > 2. At the highest *TSR* tested, σ($C_P$) and σ($C_T$) corresponding to the elevated *Ti* case were 4.5 times and 2.7 times (respectively) the corresponding values in the quasi-laminar flow. The observed σ($C_T$) was comparable to the observations reported by Mycek et al. [31]. However, the observed σ($C_P$) seemed to be noticeably larger for the tested tidal turbine model and is conjectured to be a consequence of the characteristics of the elevated *Ti* tested in the current work. Such magnitudes of load fluctuations are significant and may result in a reduction of the operational life of tidal turbines in natural marine environments[30, 31].

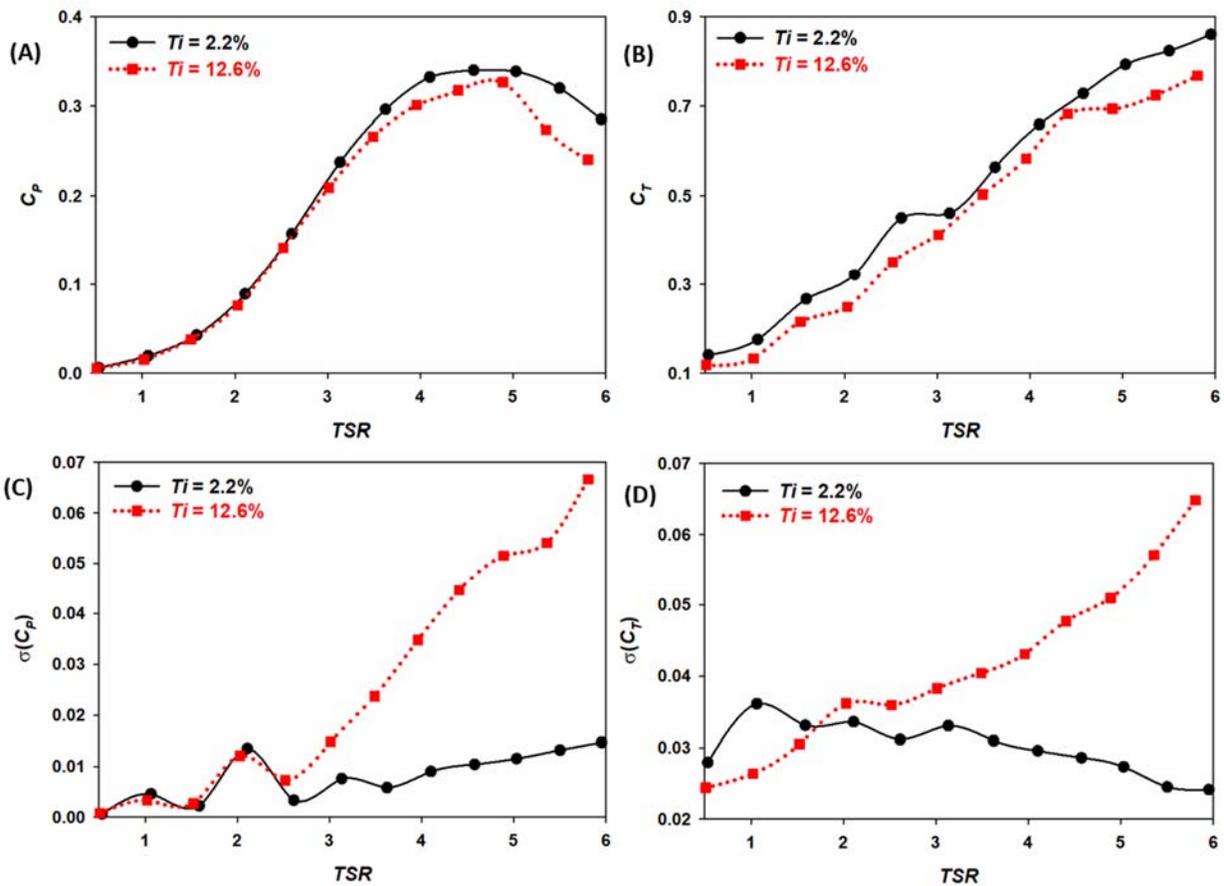

Figure 7. (A) Time-averaged power coefficient, $C_P$, (B) time-averaged thrust coefficient, $C_T$, (C) standard deviation of the power coefficient, $\sigma(C_P)$ and (D) standard deviation of the thrust coefficient, $\sigma(C_T)$ measured at the two *Ti* levels.

### 3.2 Wake velocity deficit and Swirl characteristics

All wake measurements reported in this paper were performed at the *TSR* (~5) corresponding to peak performance. The effects of inflow turbulence on rotor wake characteristics were explored in terms of the wake velocity deficit $U^*$ defined by the equation

$$U^* = \frac{U_\infty - U}{U_\infty} \tag{9}$$

where $U$ is the time-averaged velocity at that specific location and $U_\infty$ is the area averaged inflow velocity corresponding to each *Ti* experiment. $U^*$ contours at downstream locations $X/D$=0, 0.5, 1, 2 and 4, are presented in figure 8(A)-(B) for both values of *Ti*. The generation and propagation of a slow-moving, rotating wake downstream of the rotor can be observed in the figure. To facilitate a more quantifiable discussion of wake evolution, $U^*$, measured along mid-depth ($Z$=0) is plotted in figure 8(C)-(F). At $X/D = 0.5$, the velocity deficit pattern in both $Ti$ cases resembles a top-hat like profile representative of an undisturbed near-wake[31]. The most conspicuous difference between the two plotted profiles is the level of bypass flow acceleration observed beyond a radial distance of $0.6D$ from the rotor axis. The acceleration in the elevated $Ti$ was ~65%

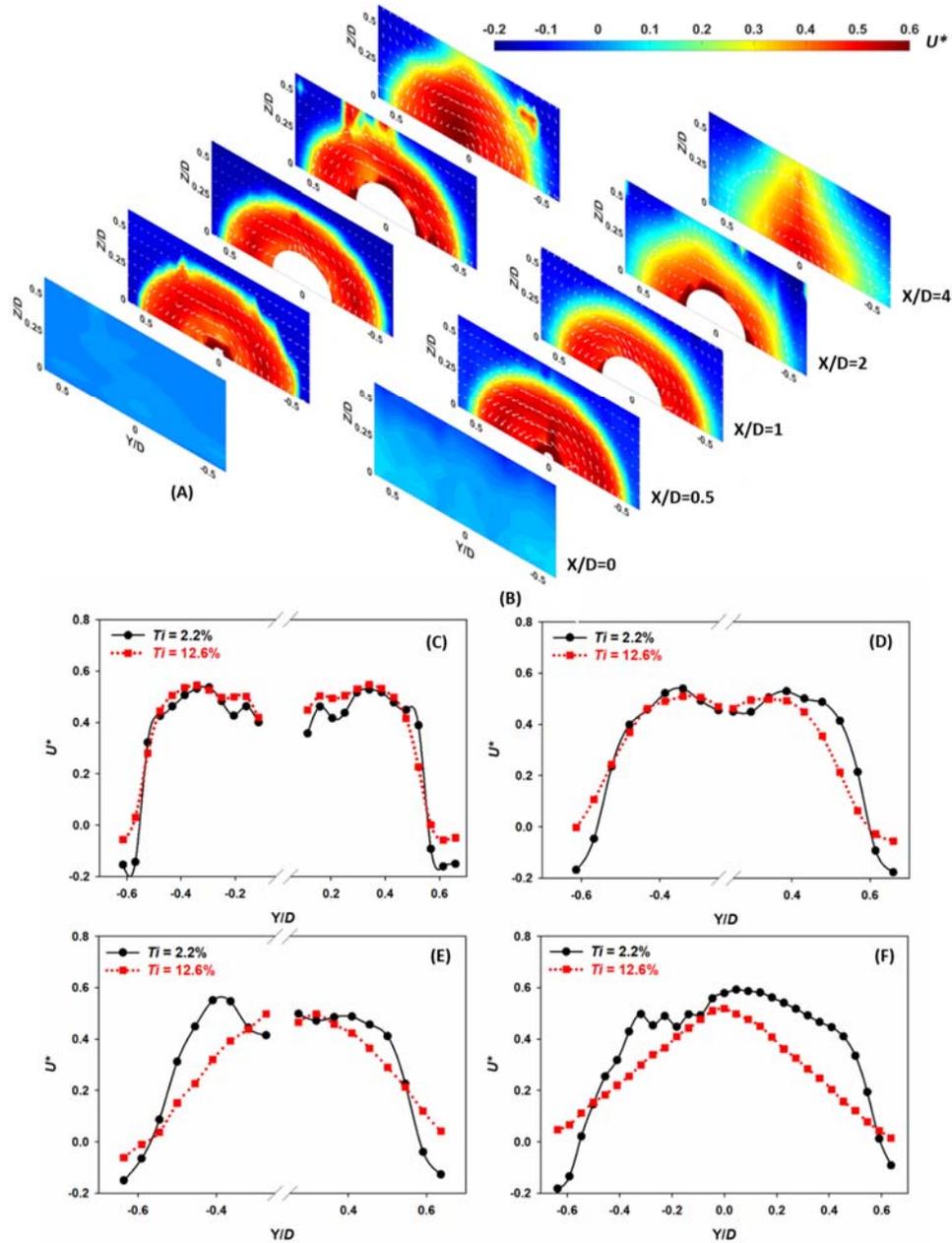

Figure 8. $U^*$ contours for (A) quasi-laminar flow case, (B) elevated $Ti$ case; mid-depth $U^*$ profiles at downstream locations (C) $X/D = 0.5$, (D) $X/D = 1$, (E) $X/D = 2$, (F) $X/D = 4$

lower than the acceleration in quasi-laminar flow, hinting towards a lesser blockage effect in higher ambient turbulence. The maximum velocity deficit calculated for the two cases was comparable (~ 0.55) at $X/D = 0.5$ and was found to be closer to the tip of the blades similar to the observation made by Tedds et al. at a radial distance of ~$0.35D$ from the axis of rotation[46]. A zone of comparatively faster moving flow (lower $U^*$) was noticeable close to the axis of rotation at $-0.2 < Y/D < 0.2$ at the downstream location of $X/D = 0.5$. This zone is believed to be a consequence of the lower solidity of the rotor close to the hub; in our turbine model, blades with a chord length of 0.017m are attached to the rotor bub via a 0.007m diameter circular section[47]. This local shrinkage in the frontal area provides a lesser blockage and an easier passage for the incoming flow resulting in a noticeable increase of local streamwise velocities. The tendency of the wake to recover faster in higher ambient turbulence begins to emerge at further downstream locations ($1 < X/D < 2$). The maximum velocity deficit in elevated $Ti$ was lower in comparison to the maximum velocity deficit measured in the quasi-laminar flow at this location (at $X/D=1, 2$). Similar values were also obtained at the upstream location $X/D=0.5$. In addition, an increase in shear layer width is noticeable in the more turbulent inflow; this is reflected by the more gradual rise in velocity deficit observed in the elevated $Ti$, at the downstream location $X/D=2$. Such differences in wake velocity gradient also indicate a transition from a top-hat like velocity profile (very near wake; i.e., $X/D \leq 0.5$) to a more diffused bell-shaped velocity profile (observed at $X/D = 4$).

Velocity deficit profiles at $X/D=4$ better illustrate the effect of ambient turbulence. Being almost immediately downstream of the turbine support structure, the maximum velocity deficit in the quasi-laminar flow undergoes an increase to a value of ~0.6; the corresponding $U^*$ profile evolves away from a top-hat like shape observed at $X/D=0.5$, into a Gaussian-like profile, and is suggestive of a more diffused wake due to cross-stream momentum transfer and wake re-energization[31]. Interestingly, an asymmetric character is evident at this location in the quasi-laminar flow and is expected to be an artifact of the presence of the turbine support structure. The $U^*$ profile measured in the elevated $Ti$ displays features of a considerably more diffused flow. Unlike in the quasi-laminar flow, the max velocity deficit of ~0.52 observed at this location in the elevated $Ti$, was lower than the maximum velocity deficit observed at $X/D = 0.5$; this is observed despite the presence of the turbine support structure. Such differences in wake velocity profiles for the elevated $Ti$ case is an outcome of the enhanced turbulent mixing and faster velocity recovery that considerably enhances the cross-stream momentum diffusion in the near wake region [23, 31].

In addition to the $U^*$ contours, figure 8 (A)-(B) also depicts the planar vectors that illustrate wake rotation. A turbine wake is an example of swirl flows that undergo simultaneous axial and rotational motion[6]. The direction of wake rotation (clockwise) was found to be opposite to that of turbine rotation. The difference in the magnitude of rotational velocities (represented by the difference in length of the planar vector) is noticeable with downstream evolution and the presence of elevated levels of inflow turbulence. To explore characteristics suggestive of momentum distribution in the wake, the swirl number ($S$) (defined below) is estimated at the different downstream locations, and turbulent inflow conditions tested[6].

$$S = \frac{G_\phi}{G_x}, \text{ where } G_\phi = \int_0^R (wr)\rho u 2\pi r dr \text{ and } G_x = \int_0^R u \rho u 2\pi r dr \qquad (10)$$

$G_\phi$ is the axial flux of angular momentum and $G_x$ is the axial flux of linear momentum. A swirl flow is considered very weak if $S \leq 0.2$, weak if $0.2 < S \leq 0.5$ and strong for $S > 0.5$[6]. Figure 9 plots $S$ calculated at the two free-stream conditions as a function of downstream distance ($X/D$). In the quasi-laminar flow, the swirling motion within the wake was found to remain fairly undisturbed between $0.5 < X/D < 2$, with a maximum value of $S = 0.14$ observed at $X/D=1$. Owing to the disturbances induced by the turbine support structure, a drop in $S$ to 0.09 was measured at $X/D=4$.

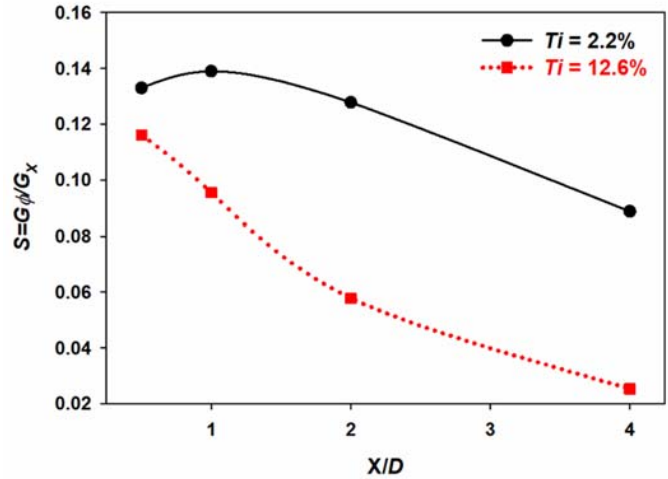

Figure 9. Swirl numbers as a function of downstream distance (*X*/*D*) for different inflow condition levels.

The wake swirl was noticeably affected in the elevated *Ti*. The maximum value of *S* calculated in elevated *Ti* was ~0.12 at *X*/*D*=0.5; a value 12% lower than the max *S* in the quasi-laminar flow. In addition, in elevated *Ti*, *S* was found to drop swiftly with downstream distance and reached a value of *S*=0.02 at *X*/D = 4. The enhanced cross stream diffusion characteristic of turbulent flow was observed to not only augment the axial flux of linear momentum but also hinder the axial flux of angular momentum. As a result, the swirl number, being a ratio of the two quantities, decreased considerably in the elevated *Ti*.

### 3.3 Wake Turbulence Characteristics

Contours of turbulence intensity in the wake of the turbine are plotted in figure 10(A)-(B). Turbulence generation downstream of a rotor can be seen to be primarily initiated in the high shear regions close to the tip and root of the rotor blades. Two sets of peaks are discernible in the mid-depth (Z=0) *Ti* profiles shown in figure 10(C) for *X*/*D*=0.5. The two outer peaks are located right at the rotor tip and signify the production of turbulence due to the interaction between the slow-moving wake and the fast-moving bypass flow. Such high levels of turbulence near the rotor tip

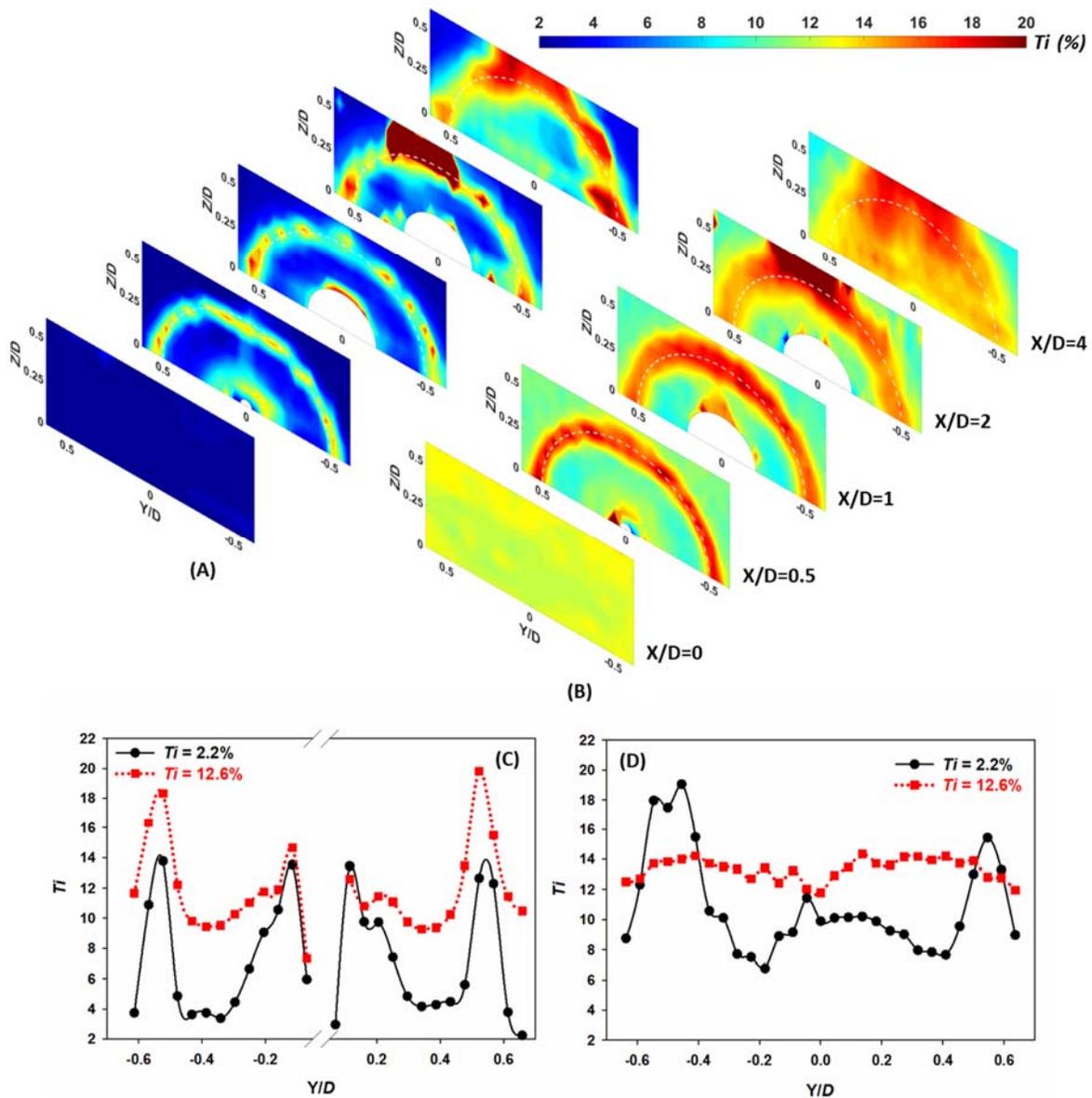

Figure 10. *Ti* contours for (A) quasi-laminar flow case, (B) elevated *Ti* case; mid-depth $Ti_X$ profiles at downstream locations (C) X/*D* = 0.5, (D) X/*D* = 4

have been reported by several studies [31, 46, 48]. In addition to the outer peak, the two inner peaks are observed in the current work and represent the turbulence generated due to the interaction between wake and the local flow acceleration resulting from the low rotor solidity close to the hub. Both inflow conditions have comparable inner peak magnitudes of *Ti* ~ 13-14%. The outer peaks, however, are noticeably different. In the quasi-laminar flow, the magnitudes of the outer peaks are comparable to the inner peaks, whereas, in elevated $Ti$, the outer peaks reach values in the range $Ti \sim 18\text{-}20\%$. Furthermore, the turbulence intensities in the annular region between the two sets of peaks were considerably different. In the quasi-laminar flow, this region had $Ti \sim 4\text{-}5\%$, whereas, at elevated $Ti$, the same region had almost double $Ti$ values in the range 9-10%. Figure 10(D) plots the mid-depth $Ti$ profiles at $X/D=4$. With the downstream evolution of the wake, the inner peaks disappear, and the $Ti$ values in the annular region increase. From figure 10(D) it can be seen that the quasi-laminar flow case continues to display the outer peaks close to the blade tip, with $Ti$ reaching values as high as 19%; $Ti$ observed in the region within the blade tips, was comparatively lower in the 7-11% range. The elevated $Ti$ case, on the other hand, displays a more diffused profile with $Ti$ bounded within the 11-14% range. The observed increase in the maximum value of $Ti$ with downstream distance (in the near wake) in the quasi-laminar flow agrees with findings reported in Mycek et al.[31], which state that, for low levels of ambient turbulence, the turbulent mixing intensification undergoes a delay and intensifies upon reaching downstream distances $\sim 5\text{--}7D$. The mid-depth Reynolds ($R_{XY}$) stress component at downstream locations, $X/D=0.5,4$ is plotted in figure 11.

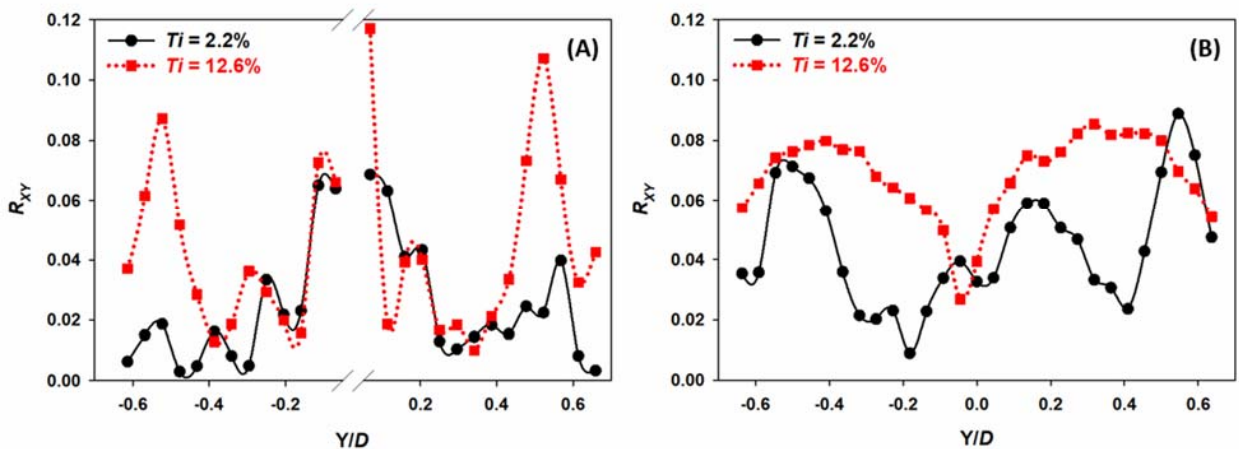

Figure 11: Mid-depth $R_{XY}$ profiles at (A) X/D=0.5, (B) X/D=4

In all cases, a correlation between the magnitudes of $R_{XY}$ and $Ti$ was noticeable. Regions with high $Ti$ also had high $R_{XY}$ that is indicative of enhanced mixing in the region. At $X/D=0.5$, the elevated $Ti$ case had considerably larger $R_{XY}$ values in regions close to the blade tip and hub; the maximum $R_{XY}$ was ~ 2 times larger than the corresponding value in the quasi-laminar flow. At a downstream location of $X/D=4$, the maximum $R_{XY}$ calculated in the two inflow conditions were found to be comparable to each other. However, the $R_{XY}$ profiles were notably different, as in figure 11(B). In the elevated $Ti$ case, significantly larger $R_{XY}$ values (2 to 3 times larger) were observed over a larger proportion of wake width. A similar increase in $R_{XY}$ (~1.5 times) with inflow turbulence intensity was also reported by Mycek et al. [31]. The elevated levels of Reynolds stress induced in the wake in a more turbulent ambient flow better reflect the enhanced cross stream momentum diffusion and a subsequent faster wake recovery.

### 3.4 Integral Length Scales

Figure 12 (A)-(B) illustrates the evolution of integral length scales ($L$, see equation 6 for a definition) in the wake of the turbine rotor. It was observed that the rotor essentially slices the integral scales in the inflow. The average value of $L$ in the quasi-laminar and elevated $Ti$ cases were estimated to be 0.22m and 0.11m, respectively. At $X/D=0.5$ downstream of the turbine, (see figure 12(C)), the value of $L$ calculated in the quasi-laminar flow dropped to ~0.006m, whereas, $L$ calculated in the elevated $Ti$ case was observed to span the ~ 0.06m–0.005m range. A near uniform $L$ distribution across the wake width was noticeable in the quasi-laminar flow; the same however was not true for the case at elevated $Ti$. In the higher ambient turbulence, $L$ calculated in regions close to the blade tip and hub were comparatively smaller as can be seen in figure 12 (C). With the downstream propagation of wake, the integral length scales gradually begin to grow in magnitude. By the downstream location of $X/D=4$ (figure 12 (D)), $L$ corresponding to the quasi-laminar flow

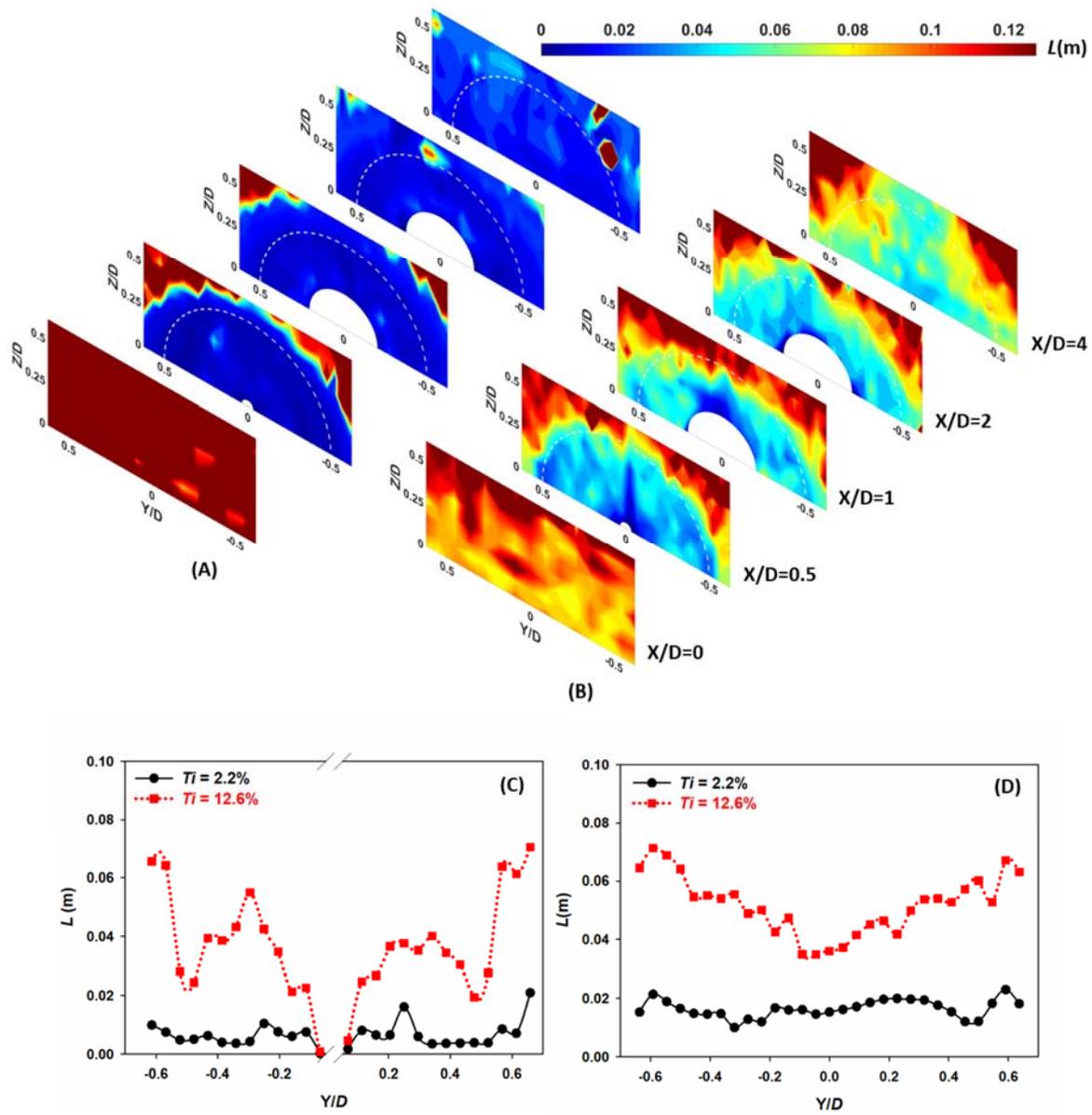

Figure 12. *L* contours for (A) quasi-laminar flow case, (B) elevated *Ti* case; mid-depth *L* profiles at downstream locations (C) X/*D* = 0.5, (D) X/*D* = 4

had grown near uniformly to an average size of ~0.016m (see figure 12(D)). The *L* profile observed at *X/D*=4 in the elevated *Ti* had become more uniform than the profile upstream at *X/D*=0.5; *L* varied between a value of 0.07m - 0.035m and is considerably larger (~ > 2 times) than the integral length scales in the quasi-laminar flow.

## 3.5 Wake Anisotropy Characteristics

The evolution of wake anisotropy downstream of the rotor is discussed next. The anisotropy ratios, $I_{XY}$, measured across the wake at two downstream locations, $X/D = 0.5, 4$ are plotted in figure 13. Immediately downstream of the rotor at $X/D = 0.5$, $I_{XY}$ across the wake width in the quasi-laminar flow is seen to fluctuate noticeably in the $0.2 < I_{XY} < 1.5$ range. The oscillations were found to be more dramatic (had larger amplitudes) in the region close to the turbine axis. A similar oscillation in the range $0.4 < I_{XY} < 2$, close to the turbine axis at $X/D = 0.5$ was noticeable even at the elevated $Ti$, however, the oscillations in wake regions away from the turbine axis were comparatively lesser.

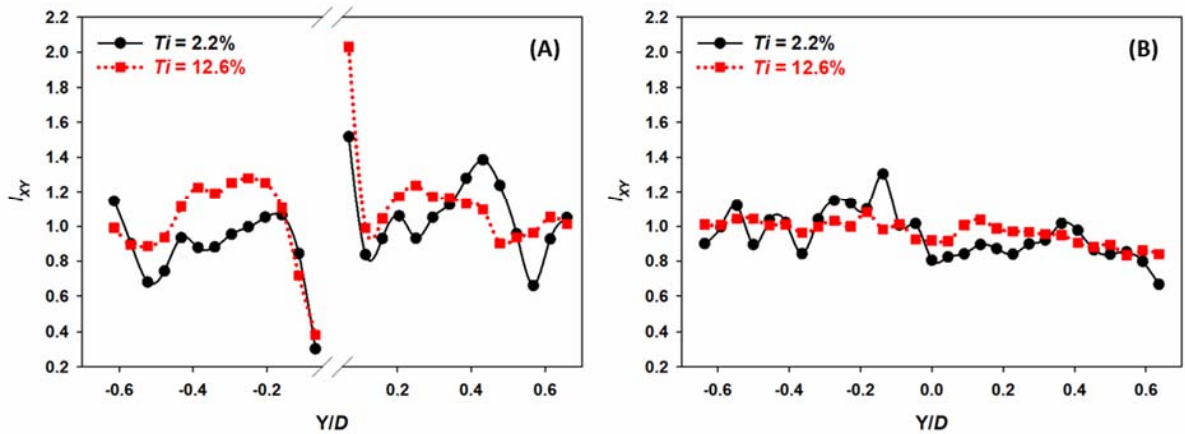

Figure 13. Mid-depth $I_{XY}$ profiles at (A) X/D=0.5, (B) X/D=4

Further downstream at $X/D = 4$, oscillations of $I_{XY}$ in the quasi-laminar flow, though noticeable, decrease in magnitude, varying between $0.7 < I_{XY} < 1.3$, while still reflecting considerable levels of anisotropic behavior (see figure 13(B)). On the other hand, $I_{XY}$ oscillations at $X/D = 4$ in the elevated $Ti$ case nearly vanish, generating a near-isotropic flow, with anisotropy ratio varying in the range $0.9 < I_{XY} < 1.1$. This implies that the enhanced stresses and cross-stream momentum diffusion induced in the wake in a more turbulent ambient flow lead to a quicker redistribution of turbulence kinetic energy among flow velocity components encouraging isotropic behavior in the wake. Sarlak *et al.*[49] used LES simulations to explore the effects of blockage on the performance and wake characteristics of a wind turbine model in a wind tunnel. They pointed out notable changes in the mean velocity of the wake due to blockage, however, were not able to identify a significant impact on the wake mixing rate. Wake anisotropy could vary with blockage ratio; the current experiments are at a blockage ratio of 16% and no inference can be drawn about the influence of blockage on anisotropy.

### 3.6 Wake Recovery Calculations

The flow energy available, *E*, in the rotor area at a particular streamwise location can be estimated through the following summation

$$E = \frac{1}{2} \rho \sum_{i=1}^{N} A_i V_i^3 \tag{10}$$

where, *N* is the total number of area elements within the frontal turbine area, and $A_i$ and $V_i$ represent the area and averaged velocity within the *i*th element. The number of elements, *N*, is taken to be equal to the number of flow measurement locations (within the rotor plane) at a given streamwise location and $A_i$ the product of the distances between adjacent horizontal and vertical locations; $A_i$ is assumed to be centered at the *i*th measurement location. To better quantify the aspect of wake recovery, the energy available within the turbine wake at the different downstream locations studied is evaluated and compared to the energy available in the inflow. Figure 14 shows the energy available/energy recovery at downstream locations *X/D*=0.5 to 4 as a percentage of the energy available at *X/D*=0. For both the inflow cases, the drop in flow energy across the rotor plane is significant, ~83%. The gradual increase in energy levels with downstream propagation reflects the process of wake re-energization through momentum diffusion from the undisturbed bypass flow. In the quasi-laminar inflow case, the energy recovery estimated at X/D=4 is ~19%. A drop in energy recovery is noticeable at X/D = 1, 2 in the quasi-laminar inflow; this can be attributed to the presence of the turbine nacelle at these two locations, which blocks off a portion of the wake cross-sectional area to the flow thereby resulting in reduced flow energy estimates at these two downstream locations. The wake recovery estimated in the elevated $Ti$ case is considerably quicker. Unlike in the quasi-laminar inflow, no drop in energy recovery was observable at the downstream locations X/D = 1, 2 and the energy recovery estimated at X/D = 4 was ~37%, almost twice the corresponding value in the quasi-laminar inflow. The presented results clearly illustrate the ability of elevated ambient turbulence to quicken the process of wake re-energization. This aspect of ambient turbulence is highly advantageous from a tidal farm standpoint as it would facilitate a denser configuration of turbine units within a farm, resulting in higher energy yield per unit area and thereby reduced costs.

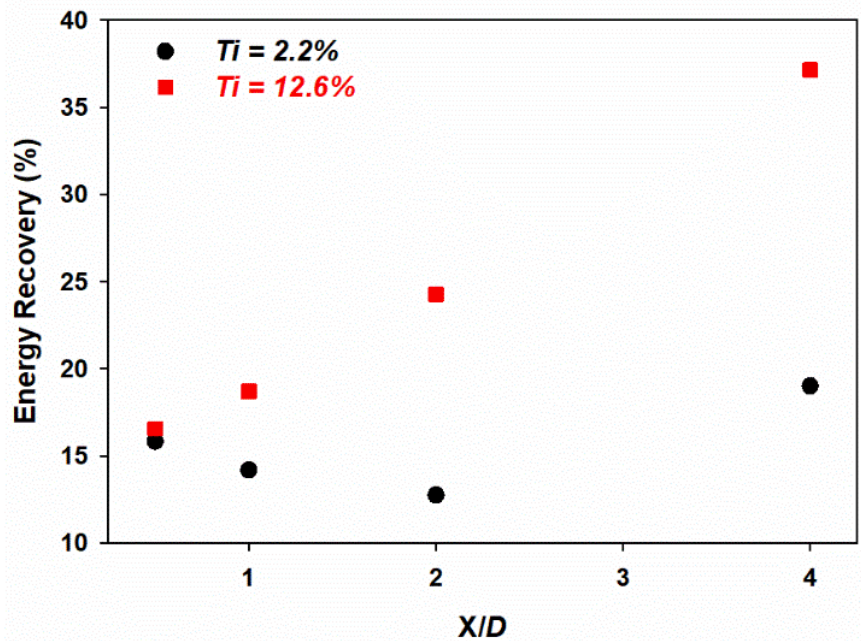

Figure 14. Wake energy recovery estimations

**3.7 Wake Periodicities and their Propagation.**

Another important concern from a tidal farm standpoint is the extent of propagation of coherent flow structures shed from upstream turbine units. In addition to inflow turbulence, periodic interaction with such structures (tip vortices for example) can be very detrimental to the operational life of turbines as they can significantly accelerate fatigue damage to downstream turbine units, necessitating additional operational/maintenance procedures and hence costs. Figure 15 shows the frequency spectra in the wake of the turbine at the radial location $Y/D = 0.54$, for two downstream locations $X/D=0.5$ and 4. From the figure, it can be seen that in the quasi-laminar flow, spectral peaks can be identified at the rotational frequency of rotor, $F_R$ (5 Hz), and its harmonics. At the downstream location $X/D=0.5$, the energy contained in the blade passing frequency $F_B$ ($3F_R$) clearly dominates $F_R$ and its other harmonics indicating the strong presence of blade tip vortices.

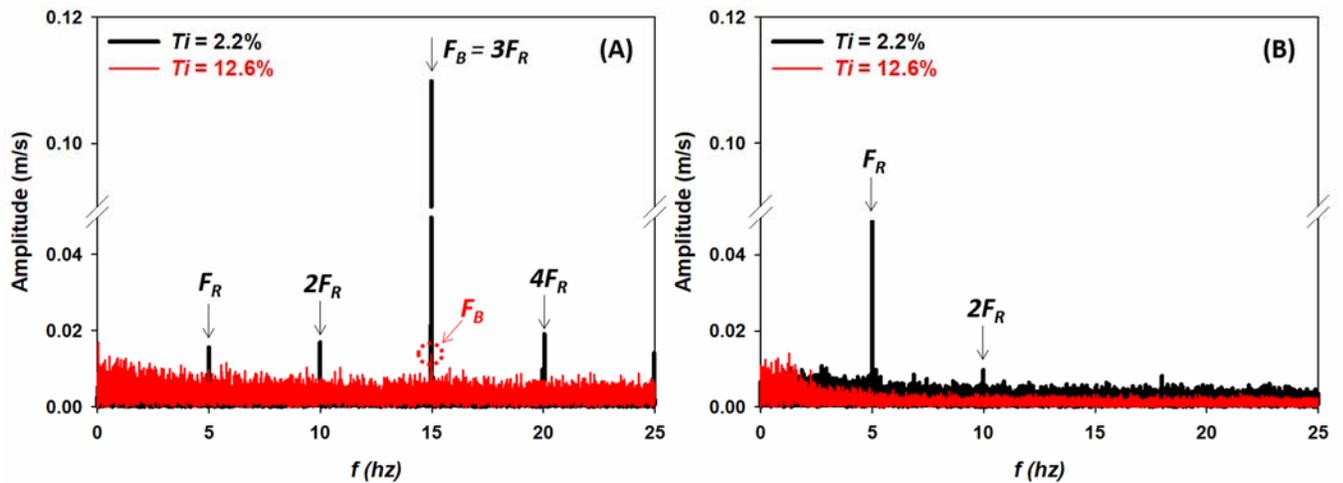

Figure 15. Frequency spectrum of the streamwise velocity component at (A) $X/D=0.5$, $Y/D=0.54$, and (B) $X/D=4$, $Y/D=0.54$

The spectral characteristics in the elevated Ti case are noticeably diffused. The energy corresponding to the rotor frequency, FR, and its harmonics are significantly lower than their counterparts in the quasi-laminar flow with FB being the only noticeable component. With downstream propagation and the consequent interaction with the free stream, the spectral energy corresponding to the dominant frequencies begins to drop. In the quasi-laminar flow, a redistribution of spectral energy is noticeable; FR is noticed to have higher spectral energy at X/D=4 than at X/D=0.5. However, its harmonics are no longer identifiable, except for 2FR. This suggests a state of persisting wake rotation with no blade tip vortices.  In the elevated Ti case, on the other hand, no periodicities are identifiable across the wake at X/D = 4, suggesting a near-complete breakdown of tip vortices and wake rotation.

### 3.8 Effects of inflow integral length scale

From the data collected in the elevated $Ti$ and elevated $Ti$-$L_D$ conditions, it was observed that both the mean and fluctuating performance quantities were insensitive to length scale effects in the tested range. The invariance in the mean quantities is expected based on findings in the literature[31]. However, the insensitivity of the fluctuating loads  suggests that the increased load fluctuations observed in a turbulent free-stream is more of a consequence of the level of turbulence intensity in the flow and not the size of structures in the flow for the range tested in the experiments reported in this paper. However, larger variations in the integral length scale (size of structures) would lead to commensurate changes in turbulence intensity and consequently could be expected to affect the load fluctuation acting on the turbine.  The variation in integral length scales generated had a considerable influence on wake statistics such as $Ti$, $L,$ and $I_{XY}$; other parameters like ($U^*$, $R_{XY}$) remained unaffected and are therefore not discussed. $Ti$ profiles across the wake of the turbine are illustrated in Figure 16(A,B); at $X/D$ = 0.5, it can be seen that $Ti$ across the wake for the elevated $Ti$-$L_D$ case is consistently larger than the corresponding values at elevated $Ti$, reaching values in the 20-23% and 26-28% ranges at the inner and outer peaks respectively. This noted increase, which is considerably larger than the differences observed between the laminar free stream and the elevated $Ti$ case is expected to be an artifact of the difference in inflow length scales in the elevated $Ti$ and elevated $Ti$-$L_D$ cases. The turbulence intensities observed in the annular region between the rotor tip and root can also be observed to be larger in the elevated $Ti$-$L_D$ case. Similar to the observations at elevated $Ti$, the turbulence intensity profile across the wake at elevated $Ti$-$L_D$ becomes more diffused with the downstream evolution; however, the variability across the wake can be noticed to remain comparatively larger. In the elevated $Ti$ case, turbulence intensity across the wake at $X/D=4$ can be observed to be uniformly restricted to values < 14%; in the elevated $Ti$-$L_D$ case on the other hand, a slightly more turbulent core region, $-0.2<Y/D<0.2$, with $Ti$ close to 16% can be identified. Figure 16 (C,D) compares the $L$ distribution across the wake in the two inflow cases. At $X/D=0.5$, it can be noticed that the variation in $L$ across the wake width follows a similar trend in both cases, although, with notably different magnitudes. $L$ estimates in the region close to the rotor hub were comparable for the two cases, however, in regions beyond the rotor hub, $L$ estimates in the elevated $Ti$-$L_D$ case were considerably larger. In the elevated $Ti$ case, integral length scales in the wake at $X/D=0.5$ were restricted to values below 0.05m, whereas, in the elevated $Ti$-$L_D$ case, $L$ estimates ranged between 0.05m – 0.12m. This observed increase in the length scales in the wake can be directly correlated to the scale difference of the incoming flow. The downstream evolution of the scales in the two cases is noticeably different. While $L$ distribution at $X/D=4$ in elevated $Ti$ becomes more uniform in the 0.03m – 0.07m range, the same in the elevated $Ti$-$L_D$ case displays a large gradient varying from values in the range 0.12m- 0.16m at wake edges to 0.03m at the wake core. The observed gradient is conjectured to be a product of two competing influences; the tendency of the larger scale ambient turbulent flow to promote the development of larger-scale structures and the tendency of the spinning rotor to slice the incoming structures. This explains the similarity of scales at the core of the wake and the divergence observed closer to the wake edges.

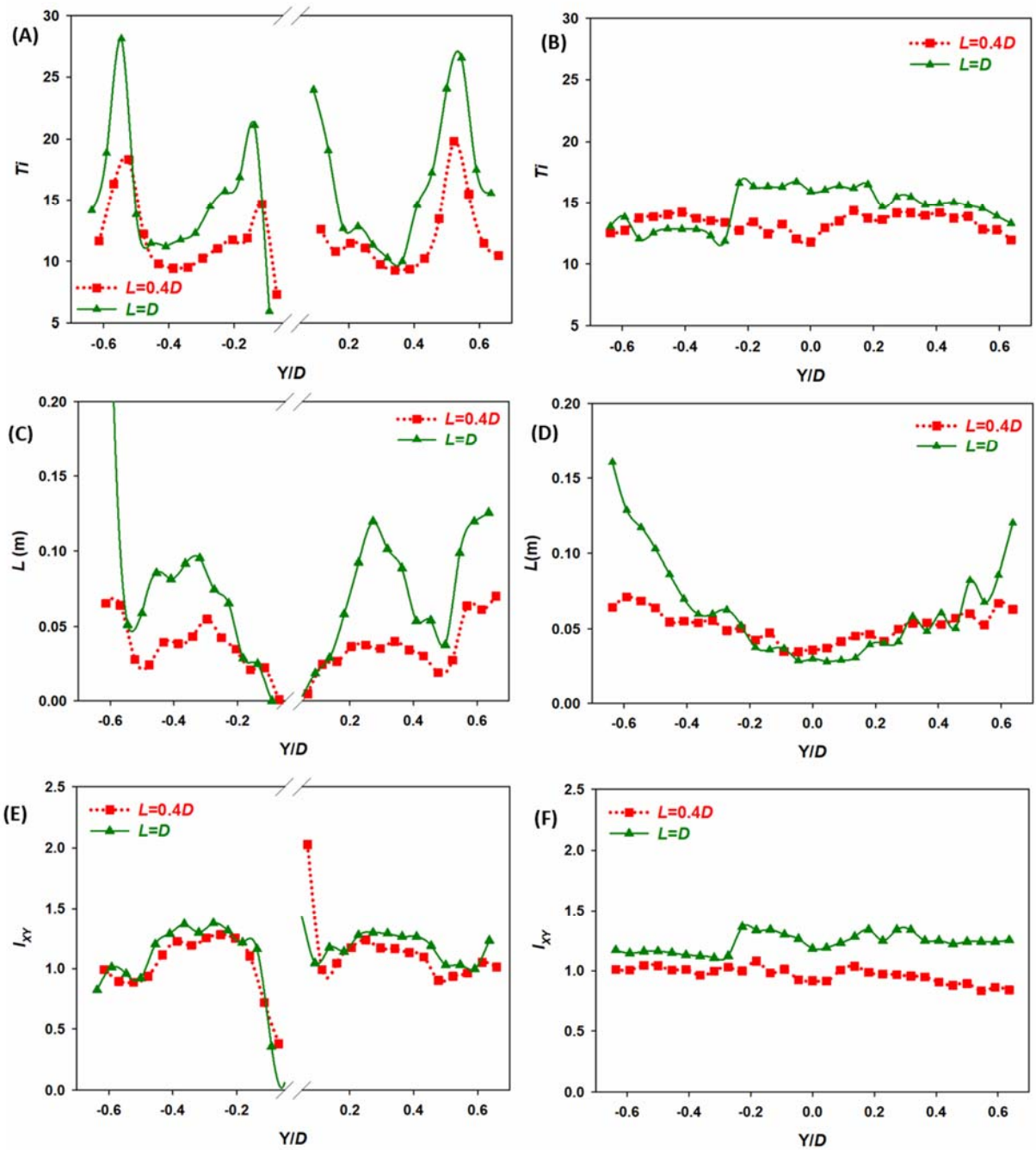

Figure 16. Effect of inflow $L$ on $Ti$, wake $L$ and $I_{XY}$ at (A),(C),(E) $X/D$=0.5 and (B),(D),(F) $X/D$=4

The anisotropy, $I_{XY}$ across the wake of the turbine, is shown in figure 18(E,F). At $X/D$=0.5 near the rotor, $I_{XY}$ distribution across the wake width can be observed to very similar in the elevated $Ti$ and elevated $Ti$-L$_D$ cases, with both being substantially anisotropic. The effect of inflow integral length scales on wake anisotropy becomes more evident with the downstream evolution of the wake. While the anisotropy levels in the wake diminish at elevated $Ti$, the elevated $Ti$-L$_D$ case leads to observably larger anisotropy levels (>20%) at $X/D$=4, reaching close to ~40% in the core of the wake.

Based on the obtained experimental data, it can be said that variations in $L$ in the range 0.4$D$ to $D$ lead to notable changes in certain wake characteristics. It leads to the existence of larger length scales/ structures in the turbine wake, thereby generating larger levels of velocity fluctuations by their interaction and break up. The larger $L$ values also lead to higher anisotropy levels in the wake, more evident at a downstream location of $X/D$=4. The 0.4$D$ to $D$ range variation in $L$ has almost no influence on turbine performance characteristics and energy recovery in the wake of the turbine. This observation further clarifies that the effects described here are dominated by the aspect of inflow turbulence intensity and not the 0.8$D$ to 0.4$D$ variation in $L$ between the quasi-laminar and elevated $Ti$ cases.

## 4. CONCLUDING REMARKS

The current experiments employ an active grid type turbulence generator to produce elevated levels of free-stream turbulence in a water tunnel facility. The effects of two important aspects of turbulent inflow, turbulence intensity ($Ti$) and integral length scale ($L$), on the performance and near-wake characteristics of a laboratory scale (1:20 ratio) tidal turbine is presented. The following conclusions can be drawn based on the measurements:

- Elevated levels of $Ti$ ($Ti$ = 12%) has a small impact on the time-averaged values of $C_P$; however, it has a marked influence on the load fluctuations acting on the turbine. The elevated $Ti$ case led to a ~10% drop in peak $C_P$ while increasing $\sigma(C_P)$ by a factor of 4.5. Such large load fluctuations have the potential to severely impact the operation and longevity of tidal turbines and their components.

- The swirl number ($S$) quantifies the distribution of momentum flux in the turbine wake. In addition to fostering the streamwise velocity component in the wake, the enhanced cross-stream momentum diffusion characteristic of higher $Ti$ was observed to also disrupt the circumferential velocities. As a result, the wake swirl numbers calculated in elevated $Ti$ were lower than the corresponding values in the quasi-laminar flow by 12% at $X/D$=0.5 to 71% at $X/D$=4.

- The streamwise turbulence intensity ($Ti$) profiles presented to highlight the production of turbulence in the high shear regions close to the blade tip and hub. The calculated Reynolds stresses also reflect the high level of stresses in the more turbulent regions. $R_{XY}$ values estimated in elevated $Ti$ were found to be ~2 times the corresponding values in the quasi-laminar flow over significant portions of the wake.

- The integral length scale contours illustrate the evolution of length scales in the wake of the rotor. Length scales calculated in the rotor wake at elevated $Ti$ were noticeably larger and less uniform compared to the length scales calculated in the quasi-laminar flow.

- The anisotropy characteristics in the wake of the turbine were also explored in the current work. Based on the different stream-wise locations analyzed, elevated $Ti$ was found to assist the isotropic evolution of turbine wake with downstream propagation.

- Elevated $Ti$ was observed to facilitate a quicker recovery of the wake velocity deficit. Based on the available energy estimations performed in the turbine wake, the flow recovery in elevated $Ti$ at $X/D$ = 4 was ~37%, nearly twice the corresponding value in the quasi-laminar flow case. This has long-ranging implications for farm-scale deployment of tidal turbines. In high ambient turbulence levels, tidal turbines can be packed more closely than the downstream spacing's typically recommended based on analytical wake models[35].

- The disruptive nature of ambient turbulence was evident in the spectral characteristics of the turbine wake as well. While the rotor rotational frequency and its harmonics were prominent in the wake frequency spectrums at the quasi-laminar flow, the blade passing frequency was barely noticeable in the elevated $Ti$ case even at $X/D$ = 0.5.

- A 0.4$D$ to $D$ variation in inflow integral length scale was found to affect the level of turbulence intensity in the rotor wake, size of structures and anisotropy characteristics; however, the variation did not show a noticeable impact on the turbine performance characteristics and the rate of wake velocity recovery.

- The authors would like to point out that, in addition to performance characteristics, the wake characteristics of the turbine could be affected by the operating $TSR$ of the rotor. Changes in $TSR$ would alter the blockage induced by the wake of the turbine and could lead to considerable changes in the interaction between the free-stream and the turbine wake and will be explored in future studies.

**ACKNOWLEDGMENTS**

The authors would like to acknowledge the financial support from the US National Science Foundation through Grant No. 1706358 (CBET-Fluid Dynamics), and from Lehigh University through the William G. Harrach and Walker Fellowships.

18. Modali, P.K., N. Kolekar, and A. Banerjee, *Performance and wake characteristics of a tidal turbine under yaw.* International Journal of Marine Energy, 2018. **1**(1).
19. Kolekar, N. and A. Banerjee, *Performance characterization and placement of a marine hydrokinetic turbine in a tidal channel under boundary proximity and blockage effects.* Applied Energy, 2015. **148**: p. 121-133.
20. Birjandi, A.H., E.L. Bibeau, V. Chatoorgoon, and A. Kumar, *Power measurement of hydrokinetic turbines with free-surface and blockage effect.* Ocean Engineering, 2013. **69**: p. 9-17.
21. Stallard, T., R. Collings, T. Feng, and J. Whelan, *Interactions between tidal turbine wakes: experimental study of a group of three-bladed rotors.* Philosophical Transactions of the Royal Society A: Mathematical, Physical and Engineering Sciences, 2013. **371**(1985): p. 20120159.
22. Lewis, M., S.P. Neill, P. Robins, M.R. Hashemi, and S. Ward, *Characteristics of the velocity profile at tidal-stream energy sites.* Renewable Energy, 2017. **114**: p. 258-272.
23. Blackmore, T., W.M.J. Batten, and A.S. Bahaj. *Influence of turbulence on the wake of a marine current turbine simulator*. in *Proc. R. Soc. A*. 2014: The Royal Society.
24. Blackmore, T., W.M. Batten, G.U. Műller, and A.S. Bahaj, *Influence of turbulence on the drag of solid discs and turbine simulators in a water current.* Experiments in Fluids, 2014. **55**(1): p. 1637.
25. Osalusi, E., J. Side, and R. Harris, *Structure of turbulent flow in EMEC's tidal energy test site.* International Communications in Heat and Mass Transfer, 2009. **36**(5): p. 422-431.
26. Thomson, J., B. Polagye, V. Durgesh, and M.C. Richmond, *Measurements of turbulence at two tidal energy sites in Puget Sound, WA.* Oceanic Engineering, IEEE Journal of, 2012. **37**(3): p. 363-374.
27. MacEnri, J., M. Reed, and T. Thiringer, *Influence of tidal parameters on SeaGen flicker performance.* Philosophical Transactions of the Royal Society of London A: Mathematical, Physical and Engineering Sciences, 2013. **371**(1985): p. 20120247.
28. Gunawan, B., V.S. Neary, and J. Colby, *Tidal energy site resource assessment in the East River tidal strait, near Roosevelt Island, New York, New York.* Renewable Energy, 2014. **71**: p. 509-517.
29. Neary, V., K. Haas, and J. Colby, *International Standards Development for Marine and Hydrokinetic Renewable Energy - Marine Energy Classification Systems*, in *US TAG TC114* 2019.
30. Blackmore, T., L.E. Myers, and A.S. Bahaj, *Effects of turbulence on tidal turbines: Implications to performance, blade loads, and condition monitoring.* International Journal of Marine Energy, 2016. **14**: p. 1-26.
31. Mycek, P., B. Gaurier, G. Germain, G. Pinon, and E. Rivoalen, *Experimental study of the turbulence intensity effects on marine current turbines behaviour. Part I: One single turbine.* Renewable Energy, 2014. **66**: p. 729-746.
32. Mycek, P., B. Gaurier, G. Germain, G. Pinon, and E. Rivoalen, *Experimental study of the turbulence intensity effects on marine current turbines behaviour. Part II: Two interacting turbines.* Renewable Energy, 2014. **68**: p. 876-892.
33. Payne, G.S., T. Stallard, R. Martinez, and T. Bruce, *Variation of loads on a three-bladed horizontal axis tidal turbine with frequency and blade position.* Journal of Fluids and Structures, 2018. **83**: p. 156-170.
40